\documentclass[%
 reprint,
 article,
 amsmath,amssymb,
 aps,
 prb,
floatfix,
final
]{revtex4-2}

\usepackage{graphicx}
\usepackage{dcolumn}
\usepackage{bm}
\usepackage[version=4]{mhchem}
\usepackage{amssymb}
\usepackage{amsmath}
\usepackage{siunitx}
\usepackage{hyperref}
\usepackage{url}
\usepackage{microtype}

\usepackage{xcolor}
\usepackage{multirow}
\usepackage{tabularx}
\usepackage{ulem}
 
\usepackage{braket}
\usepackage{gensymb}

\usepackage[left]{lineno}

\begin{document}

\title{
Ab initio screening of quantum frustrated materials with kagome and triangular geometries
}


\author{Byeong-Hyeon Jeong} 
\author{Hee Seung Kim}      
\author{SungBin Lee}
\email{sungbin@kaist.ac.kr}
\author{Myung Joon Han}
\email{mj.han@kaist.ac.kr}
\affiliation{Department of Physics, Korea Advanced Institute of Science and Technology (KAIST), Daejeon, South Korea}
\date{\today}

%
%

\begin{abstract}
Geometrical frustration is a powerful route to realize exotic phases such as quantum spin liquids. 
Despite extensive efforts, systematic searches targeting specific frustration motifs and their potential to host unconventional magnetic ground states remain rare, thus highlighting the need for a more focused and predictive materials discovery approach.
Here we present a new strategy combining high-throughput first-principles calculations, magnetic force theory, and spin Hamiltonian analysis. Starting from the $\sim$150,000 material database, we catalogue candidate materials that may host competing exchange interactions and new types of magnetic states with the focus on kagome or triangular lattices.
Our workflow not only reproduces the majority of known frustrated magnetic materials, validating our approach, but also predicts novel candidate compounds with targeted frustration profiles that have not yet been experimentally synthesized. Among these, we identify six promising new materials: one triangular lattice compound, \ce{KMgNiIO6}, and five kagome lattice compounds— \ce{Li4Fe3WO8}, \ce{Li2V3F8}, \ce{Li5VP2(O4F)2}, and \ce{Li2MgCo3O8} ($P2/m$ and $C2/m$). For each candidate, we identify detailed magnetic properties and further propose their potential magnetic ground states, revealing that some of them may host entirely new magnetic phases driven by their distinct frustration characteristics.

\end{abstract}

\maketitle


\begin{figure*} [hbt!]  
\includegraphics[width=0.95\textwidth]{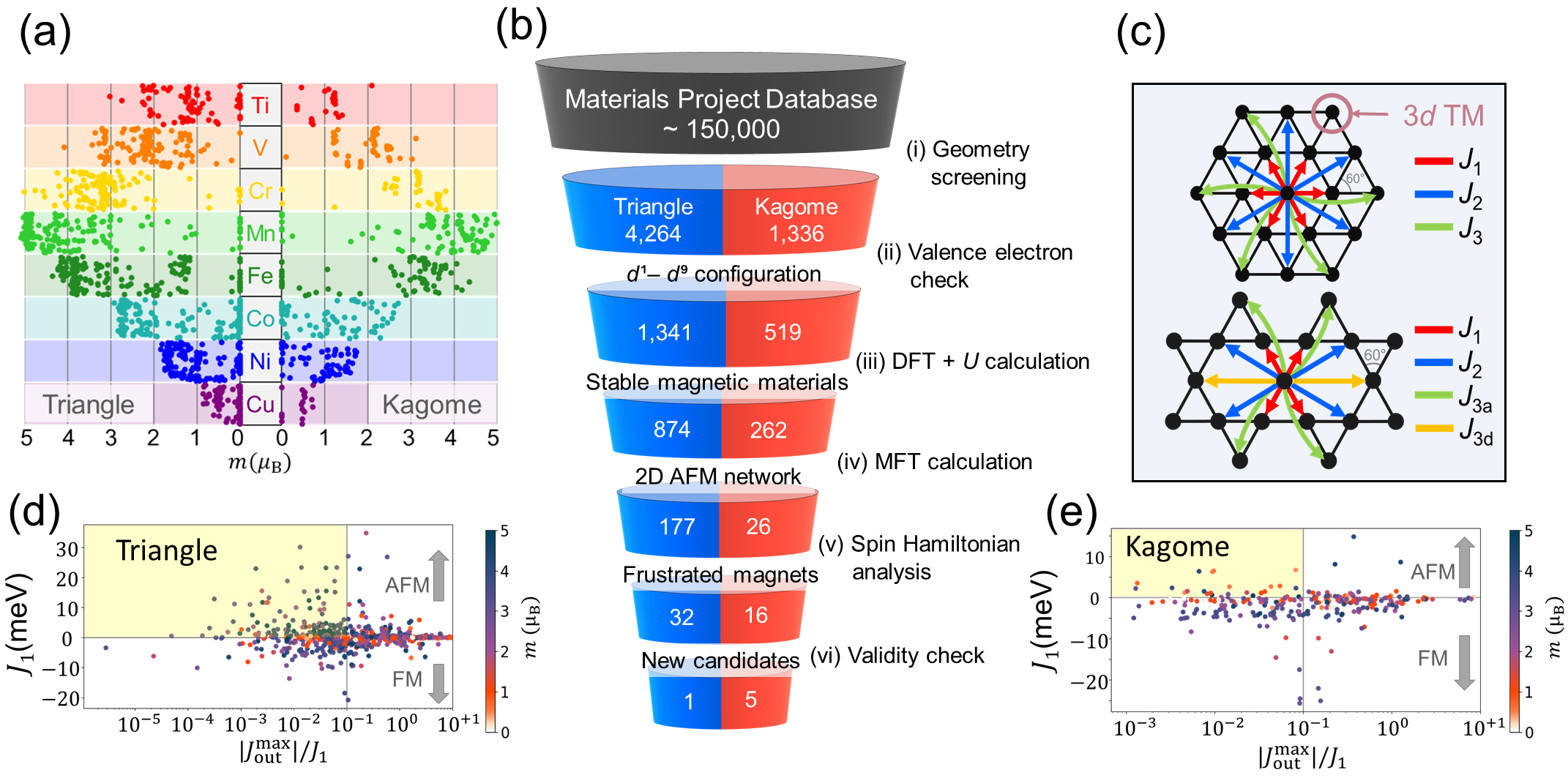}
\caption{(a) Calculated magnetic moments at TM sites for 1,653 materials yielding stable magnetic solutions. (b) Schematic overview of the multi-step high-throughput screening workflow and the corresponding results. Starting from a database of 154,713 materials, the procedure ultimately provides the catalog of frustrated magnetic materials and  identifies the six strongest candidates which have not yet been synthesized. (c) Crystal structures and associated intra-layer exchange interactions for triangular and kagome lattices. (d, e) Calculated $J_{1}$ values and their ratios with respect to the largest inter-layer coupling ($J_{\mathrm{out}}^{\mathrm{max}}$). The yellow-shaded region indicates the screening threshold $|J_{\mathrm{out}}^{\mathrm{max}}|\leq 0.1 J_{1}$ (with AFM $J_1$). Panels (d) and (e) present results for triangular and kagome systems, respectively. By convention, positive (negative) $J$ corresponds to antiferromagnetic (ferromagnetic) interactions.}
\label{fig:workflow}
\end{figure*}

\begin{figure*} [hbt!]  
\includegraphics[width=0.95\textwidth]{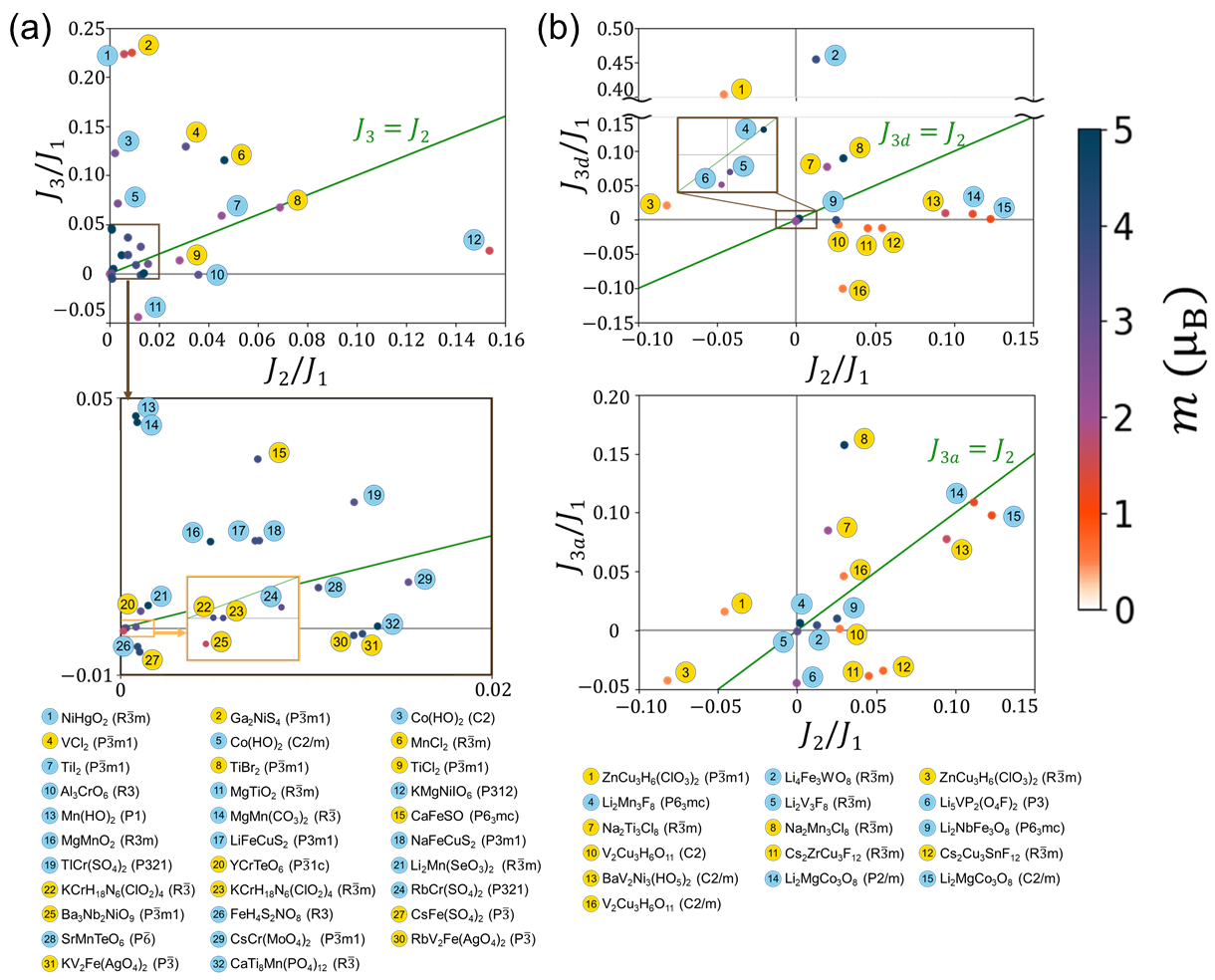}
\caption{
Catalog of (a) triangular and (b) kagome lattice compounds situated in the magnetically frustrated regime, plotted in the $J_2/J_1$--$J_3/J_1$ parameter space. Each numbered marker corresponds to a specific material, with yellow circles indicating compounds that have been experimentally synthesized and blue circles representing unsynthesized candidates.  The color scale denotes the calculated ordered magnetic moment $m$ (in $\mu_B$). 
 Green lines mark the condition $J_{3(a/d)} = J_2$, and magnified insets highlight regions with dense data distributions.
}
\label{fig:mft_results}
\end{figure*}

\section{Introduction}

Geometrical frustration arises when spins fail to simultaneously minimize their exchange energy owing to competing interactions. This phenomenon has attracted considerable interest due to its potential to host unconventional magnetic orders or quantum spin liquids (QSLs), characterized by long-range quantum entanglement and fractionalized excitations \cite{Intro_Anderson1973, Intro_Balents2016, Intro_zhou2017, Intro_wen2017, Intro_broholm2020}. Consequently, the search for frustrated magnetic materials capable of realizing such exotic ground states has remained a central focus of condensed matter physics. Over the past few decades, sustained efforts have been directed toward both the experimental discovery and the theoretical investigation.
Despite these efforts, the identification of genuine frustrated magnetic systems remains a topic of active debate \cite{Intro_Anderson1973, Intro_Balents2016, Intro_zhou2017, Intro_wen2017, Intro_broholm2020,herbertsmithite_norman2016}. It highlights the longstanding need for systematic approaches to identify frustrated magnets in a controlled and predictive manner.

To date, most theoretical and computational efforts have focused on understanding the magnetic properties of already synthesized materials. A wide range of tools have been successfully employed to characterize known systems and uncover their natures.
While these approaches have provided valuable insights, they offer limited guidance for the new materials discovery. A systematic framework and a corresponding workflow have yet to be established. Although high-throughput screenings based on the extensive material database and density functional theory (DFT) calculations have been successful for various cases \cite{mounet2018two, vergniory2019complete, zhang2019catalogue, tang2019comprehensive, xu2020high, wu_data-driven_2022, regnault_catalogue_2022, gao_high-throughput_2022,  jin_atomic_2023, li_high-throughput_2024, bjork2024two, chen_unconventional_2025}, it has been severely limited for this case of strongly correlated electron phenomena \cite{DFT2019_kagome, DFT2023_Herbertsmithite_substitution}. The targeted search for specific forms of frustration remains largely unexplored.

In this study, we establish a systematic, multi-step workflow for the discovery of novel frustrated magnetic materials, combining high-throughput screening with first-principles calculations, magnetic force theory, and spin Hamiltonian analysis. This approach enables the targeted identification of specific magnetic frustration emerging from lattice geometry and competing exchange interactions. We focus on triangular and kagome lattice compounds with partially filled $3d$ transition metal (TM) ions. Notably, our framework successfully rediscovers many well-established QSL candidates, including such kagome systems as herbertsmithite, kapellasite, Vesignieite, and Volborthite. Also identified are triangular lattice compounds, including NiGa$_2$S$_4$, Ba$_3$NiSb$_2$O$_9$ (6H-B phase), and delafossite-like AgCrSe$_2$. The successful identification of these benchmark materials validates the robustness and reliability of our approach.

Further, our workflow uncovers previously unexplored compounds predicted to exhibit strong frustration. These include a triangular lattice compound, \ce{KMgNiIO6}, and five kagome lattice candidates: \ce{Li4Fe3WO8}, \ce{Li2V3F8}, \ce{Li5VP2(O4F)2}, and two symmetry-inequivalent forms of \ce{Li2MgCo3O8} with $P2/m$ and $C2/m$ space groups. These results demonstrate the power of systematic, data-driven search in guiding the discovery of new frustrated magnets and spin liquid candidates. For each candidate, we assign the most probable ordered manifold. The newly predicted materials are strong candidates to host not only conventional coplanar orders ($q=0$) but also noncoplanar cuboc states, including the newly identified `cuboc-3' region. Our workflow provides (i) a validated, data-driven searching strategy, (ii) an experimentally actionable shortlist of frustrated magnets with quantified exchange hierarchies, and (iii) clear model-level diagnoses to guide synthesis and characterization.

\section{The high-throughput screening workflow}
Our material-search workflow consists of six steps, as shown in Fig.~\ref{fig:workflow}(b). Here we outline the main idea and the computational details are provided in Computation Details section.

\subsection{Database search and the screening for geometrical property}

We first search a well-established material database for candidate materials. In the current work, we focus on two lattice systems, namely kagome and triangular lattices, and use the Materials Project database \cite{materials_projects_database}, which contains 154,713 materials. Importantly, it includes both experimentally synthesized and unsynthesized materials. For kagome lattice systems as an example, our screening criteria include (see Fig.~\ref{fig:workflow}(c)): (a) Each kagome site should have four nearest neighbors (NNs). (b) Bond angles between two NN sites should be given by either \ang{60}, \ang{120}, or \ang{180}. Similarly, for the triangular system, each site has six NNs, and the bond angles are six of \ang{60}, six of \ang{120}, and three of \ang{180}. Since some obviously irrelevant materials can also satisfy these conditions, we perform an additional `planarity check' to ensure that all these atoms lie within a single plane. To account for unavoidable uncertainties and possible database errors, we introduce tolerances of up to \ang{4} in bond angles and up to 6\% in bond lengths \cite{struct_karigerasi2018}.


\subsection{Screening based on the valence configuration}

From the initial screening step, we obtained 4,264 triangular and 1,336 kagome materials with $3d$ TMs occupying the primary sites. By considering the oxidation states of the TM ions, we further excluded compounds with $3d^{0}$ or $3d^{10}$ electronic configurations. The valence states were determined using the tool described in Ref.~\cite{Generalic2024}. In addition, we removed materials in which multiple atomic species simultaneously possess partially filled $d$-orbitals; such systems merit further investigation in future studies. After these refinements, our screening yielded 1,341 triangular lattice and 519 kagome lattice candidate materials. 


\subsection{DFT$+U$ calculation}

The third step involves DFT$+U$ calculations for 1,860 (= 1,341 + 519) candidates since the materials of interest lie well within the strongly correlated regime. The primary goal of this DFT$+U$ step is to obtain the full-band Hamiltonian and spin-dependent charge density. However, it also serves as a further screening stage: materials whose self-consistent solutions converge to a nonmagnetic state or to a configuration with nonuniform magnetic moments were excluded. We retained only those systems that converged reliably to symmetry-preserving solutions with nonzero magnetic moments. As a result, 874 triangular and 262 kagome lattice materials were selected. The calculated magnetic moments of these compounds are presented in Fig.~\ref{fig:workflow}(a).

\subsection{MFT calculation}

We extract the magnetic coupling constants by applying the magnetic force theorem (MFT) within linear response theory \cite{liechtenstein1987, antropov1996spin, katsnelson2000first, han2004electronic}. For conventional Heisenberg spin Hamiltonian,  $H= - J_{ij} {\bf S}_i \cdot {\bf S}_j$, MFT estimates
\begin{equation}\label{Eq_Jij_all}
J_{ij}({\bf{q}}) = \frac{1}{\pi} {\rm{Im}} \int^{\epsilon_{\rm{F}}} \!\! d\epsilon ~ \int_{\rm B.Z.} \!\!\! d{\bf k}  ~
{\rm Tr}[
\mathbf{V}_{{\bf k},i} \mathbf{G}_{{\bf k},i j }   \mathbf{V}_{{\bf k+q},j} \mathbf{ G}_{{\bf k+q},ji}
],
\end{equation}
and its Fourier transform gives rise to $J_{i}$. $\mathbf{G}_{{\bf k},i j }$ is Green's function propagator, and $\mathbf{V}_{{\bf k},i}$ represents spin perturbation.
MFT enables us to avoid the multiple total-energy calculations of enlarged supercells which are not well compatible with high-throughput screening workflow. 

The calculation results are presented in Fig.~\ref{fig:workflow}(d) and (e) in terms of the averaged NN exchanges, $J_1$, and the maximum out-of-plane value ($J_{\rm out}^{\rm max}$). Before conducting further analyses in the following step, we discard the materials unless they have major AFM couplings and the well-defined 2D triangular/kagome plane: namely, (i) $J_1 >0$ and (ii) $|J_{\rm out}^{\rm max} / J_1| < 0.1$. 
The selected materials are the ones that fall into the yellow-colored regions in Fig.~\ref{fig:workflow}(d) and (e); 177 triangular and 26 kagome materials.

\subsection{Spin Hamiltonian analysis -- Screening of exchange interactions}

We next identify compounds with strongly competing exchange interactions. This step is critical: although triangular or kagome connectivity provides the geometric precondition for frustration, further-neighbor competition can substantially enhance it. Because such competition cannot be captured by a single parameter, our fifth screening stage adopts a broader, multi-metric evaluation.

For triangular lattices, previous studies have explored the enhanced frustration conditions in terms of $J_1$ and $J_2$ \cite{triangular_rubin2010spin,triangular_xu2012,Model_li2015, Model_zhu2015, Model_iqbal2016,Model_saadatmand2016,Model_ferrari2019, triangular_szasz2022phase,jiang2023nature}.
For instance, in the simple spin-$1/2$ $J_1$ Heisenberg model, the ground state is characterized by $120\degree$ magnetic order. However, the introduction of a small but positive $J_2$ ($0.08 \lesssim  J_2 \lesssim 0.16$) stabilizes a QSL phase. Motivated by this, we apply screening criteria requiring $J_2 > 0$ and $|J_{1}^{\rm avg}| > |J_{2,3}^{\rm avg}|$. To account for inevitable numerical deviations and potential database inaccuracies, we allow the standard deviation of $J_1$ values to be as large as 25\% of $J_{1}^{\rm avg}$ —a tolerance required to maintain the stability of well-established compounds. Furthermore, we verify at this stage that the converged solutions are insulating and that the total energy calculations favor antiferromagnetic states. Applying this sequence of screening conditions reduces the pool of candidate triangular materials from 177 to 32, as summarized in Fig.~\ref{fig:mft_results}(a).

 For kagome lattices, the $J_1$ Heisenberg model is broadly believed to host a spin-liquid ground state. When further-neighbor interactions are introduced, previous studies have shown that the system remains highly frustrated and does not develop long-range magnetic order, irrespective of the sign of $J_2$, provided its magnitude is small
~\cite{Model_kolley2015,discuss_gong2015,Model_liao2017,Model_changlani2018,Model_prelovvsek2020,Model_iqbal2021}. Accordingly, we do not impose any restriction on the sign of $J_2$ in our screening procedure. For total energy comparisons, we adopt the two-up–one-down spin configuration. With these criteria, we identify 16 kagome candidates, listed in Fig.~\ref{fig:mft_results}(b), which will be subjected to further analysis in the next step.

\subsection{The final selection and the validity check}

Among the 32 triangular systems, 13 have already been synthesized experimentally, including well-established compounds that reinforce the validity of our screening workflow. For the 19 unsynthesized candidates, we analyze their magnetic coupling profiles in detail, guided by previous model studies \cite{triangular_rubin2010spin,triangular_xu2012,Model_li2015, Model_zhu2015, Model_iqbal2016, Model_saadatmand2016, Model_ferrari2019, triangular_szasz2022phase, jiang2023nature}. In particular, we focus on the criterion $J_2 \gtrsim 0.08 $, where the conventional 120$\degree$ magnetic order is expected to be destabilized. Among the unsynthesized, \ce{KMgNiIO6} ($R\bar{3}m$, \#9) emerges as the only promising candidate for which we perform a series of additional check calculations considering the known limitations of conventional DFT$+U$ methodology (see Computation Details section and Appendix). This material is finally selected.

As for kagome systems, 9 out of the 16 candidates have already been synthesized experimentally. This list again includes well-known QSL candidates such as Herbertsmithite and Volborthite, further supporting the robustness of our approach. For the 7 unsynthesized kagome materials, we carry out detailed investigations. 
The same validation check procedure with different DFT functional choices and interaction parameters excludes two: \ce{Li2Mn3F8} ($P6_{3}mc$, \#4) and \ce{Li2Nb3Fe3O8} ($P6_{3}mc$, \#9). 

For the final list of six new materials (1 triangular + 5 kagome; see Table~\ref{dft_table}), further discussion will be discussed in the following section.

\begin{table*} [htb!]
        \setlength{\tabcolsep}{8pt}
	\begin{tabular}{crrrrrrr} \hline\hline 
					&	\ce{KMg\textbf{Ni}IO6}	&	\ce{Li4\textbf{Fe}3WO8}	&	\ce{Li2\textbf{V}3F8}	&	\ce{Li5\textbf{V}P2(O4F)2}	&	\ce{Li2Mg\textbf{Co}3O8}	&		\ce{Li2Mg\textbf{Co}3O8}    &    \\		
					&	mp-2222843	&	mp-756528	&	mp-778241	&	mp-758318	&	mp-1277613	&	mp-1177928	\\	\hline	
		Lattice	    &	Triangle (\#12)	&	kagome (\#2)	&	kagome (\#5)	&	kagome (\#6)	&	kagome (\#14)	&	kagome (\#15)	\\		
		Space group	&	$P312$	&	$R\bar{3}m$	&	$R\bar{3}m$	&	$P3$	&	$P2/m$	&	$C2/m$	\\		
		$J_1$ (meV)	&	0.079	&	0.653	&	0.922	&	0.036	&	2.592	&	2.737	\\		
		$J_2/J_1$	&	0.154	&	0.013	&	0.000	&	0.000	&	0.112	&	0.123	\\		
		$J_{3(a)}/J_1$	&	0.023	&	0.004	&	$-$0.002	&	$-$0.045	&	0.114	&	0.098	\\		
		$J_{3d}/J_1$	&	-	&	0.454	&	0.000	&	$-$0.002	&	$-$0.001	&	0.001	\\		
		$J^{\rm{out}}/J_1$	&	0.008	&	0.003	&	$-$0.001	&	$-$0.063	&	0.031	&	0.001	\\		
		$m$ ($\mu_B$/TM)	&	1.6	&	3.80	&	3.11	&	2.21	&	1.28	&	1.19	 \\ \hline
        \text{Predicted ground state} & Stripy/$120\degree$ & cuboc-1 & CSL & valence-bond &  $q=0$/cuboc-3 &  $q=0$/cuboc-3 \\
        \hline\hline
	\end{tabular}\label{Table_Dudarev} 
	\caption{Summary of six newly identified candidate magnetic materials, their crystallographic lattices, exchange-interaction parameters, and the theoretically predicted magnetic ground states. The first row lists the corresponding Materials Project identifiers (mp-IDs). The dominant exchange coupling $J_1$ (in meV) and its ratios with further-neighbor and interlayer couplings ($J_2/J_1$, $J_{3(a)}/J_1$, $J_{3d}/J_1$, and $J_{\mathrm{out}}/J_1$) are obtained from first principles calculation and magnetic force theory. For kagome systems, $J_{3(a)}$ and $J_{3d}$ correspond to the two distinct in-plane third-neighbor exchanges respectively, as marked in Fig.\ref{fig:workflow}(c). The ordered magnetic moment per transition-metal (TM) ion, $m$, is listed in units of $\mu_B$. 
    The last row indicates the predicted ground-state spin configuration, including noncollinear orders such as stripy and 120$^{\circ}$ states in the triangular lattice, as well as noncoplanar cuboc-1 and cuboc-3 orders and classical spin liquid (CSL) states in the kagome case. The TM species carrying local magnetic moments are indicated in bold.
    }
\label{dft_table}
\end{table*}

\section{Results}

\subsection{Catalog of triangular materials}

Following the workflow described above, the selected 32 triangular materials are expected to host well-defined triangular spin lattices localized on the transition-metal ions, exhibit insulating gaps, and feature two dominant competing antiferromagnetic interactions ($J_{1,2} > 0$).

Among the 13 synthesized compounds highlighted in Fig.~\ref{fig:mft_results}(a) by yellow labels, 120° AFM order has been reported for  
\ce{VCl2} ($P\bar{3}m1$, \#4),
\ce{TiCl2} ($R\bar{3}m$, \#9),
\ce{CaFeSO} ($P\bar{3}$, \#15), 
\ce{Ba3Nb2NiO9} ($P\bar{3}m1$, \#25),
\ce{CsFe(SO4)2} ($P\bar{3}$, \#27),
\ce{RbV2Fe(AgO4)2} ($R3m$, \#30), and
\ce{KV2Fe(AgO4)2} ($P\bar{3}$, \#31) 
\cite{reynolds1999crystal, inami2007neutron, hwang2012successive, amuneke2014experimental, lewis1962380}.
These compounds also carry sizable ordered moments (as indicated by the data-point color scale). By contrast, \ce{Ga2NiS4} ($P\bar{3}m1$, \#2) is highly frustrated with only short-range correlations; although its exact ground state and the values of its competing exchange interactions remain debated, there seems to be broad consensus that third-neighbor coupling dominates, with much weaker second-neighbor exchanges~\cite{mat_nambu2011,mazin2007ab,stock2010neutron,nakatsuji2010novel,zhang2014monte, pradines2018study}. This is consistent with our results (see Fig.~\ref{fig:mft_results}(a)).

For several other cases, the available experimental information remains insufficient. For instance, no magnetic measurements have been reported for \ce{TiBr2} ($P\bar{3}m1$, \#8), \ce{YCrTeO6} ($P\bar{3}1c$, \#20), \ce{KCrH18N6(ClO2)4} ($R\bar{3}$, \#22), and \ce{KCrH18N6(ClO2)4} ($R\bar{3}m$, \#23). In the case of \ce{MnCl2} ($R\bar{3}m$, \#6), previously reported experimental results appear inconsistent, likely due to its chemical instability and possible intercalation effects~\cite{wiesler1995determination,mcguire2017crystal}.

Importantly, the result of synthesized materials reinforces the validity of our workflow. This observation is consistent with previous theoretical predictions and indicates that our approach effectively identifies frustrated magnetic systems. We therefore proceed to the list of unsynthesized candidates, indicated in blue in Fig.~\ref{fig:mft_results}(a). Most of these materials are well below the critical boundary, i.e., $J_2/J_1 \lesssim 0.08$, which signals the destabilization of 120$^\circ$ magnetic order \cite{triangular_xu2012,Model_li2015,Model_zhu2015,Model_iqbal2016,Model_saadatmand2016,Model_ferrari2019,jiang2023nature}. The only exception is \ce{KMgNiIO6} ($P312$, \#12).\\[2mm]

{\bf \text{\ce{KMgNiIO6}} ($P312$, \#12) --- }
The detailed descriptions of its electronic structure, crystal symmetry and magnetic couplings are given in Appendix (Fig.~4 and Table~IV). Importantly, this material lies close to the phase boundary between stripy and 120$^\circ$ order in spin-1 $J_1-J_2$ Heisenberg model \cite{triangular_rubin2010spin, triangular_xu2012, triangular_szasz2022phase}. Our MFT calculations yield a ratio of $J_2/J_1 = 0.154$, with a small enough $J_3$, as shown in Fig.~2(a). Being near the phase boundary, with $J_2/J_1 \approx 0.13$ (see also Appendix, Table~IV), the system may host an intriguing magnetic phase, such as a stripy order beyond the conventional $120^{\circ}$ state
\footnote{Given that it carries $S \!\sim\! 1$ moments, a more rigorous analysis is warranted, particularly beyond the bilinear Heisenberg framework. In this regime, higher-order spin interactions—such as biquadratic exchange—can play a significant role~\cite{PhysRevLett.97.087205,PhysRevB.90.144409,10.21468/SciPostPhys.3.4.030,PhysRevB.97.245146}.}.

\subsection{Catalog of kagome materials}

The final list of 16 kagome compounds exhibit insulating gaps and localized spin moments at the transition-metal sites with dominant antiferromagnetic couplings ($J_1 > 0$). They are presented in Fig.~\ref{fig:mft_results}(b), where the 9 experimentally synthesized and 7 unsynthesized materials are distinguished.

First, we examine whether the proposed workflow reproduces the well-established kagome compounds known to host frustrated magnetism, including prominent QSL candidates. Previous theoretical studies have indicated that the characteristic regime of interest satisfies $J_2/J_1 \ll 1$ and $J_{3a}, J_{3d} \ll J_1$~\cite{Model_kolley2015,Model_liao2017,Model_changlani2018,Model_prelovvsek2020,Model_iqbal2021}. Encouragingly, our screening procedure successfully recovers several benchmark systems within this regime. For example, Herbertsmithite (\ce{ZnCu3H6(ClO3)2}, $R\bar{3}m$, \#3)~\cite{mat_Herbertsmithite2010,mat_Herbertsmithite2012,mat_Herbertsmithite2015} and Volborthite (\ce{V2Cu3H6O11}, $C2/m$, \#16)~\cite{mat_Volborthite2001,mat_Volborthite2019} both exhibit second- and third-neighbor exchanges ($J_2$, $J_{3a}$, $J_{3d}$) smaller than 10\% of $J_1$. A structurally related phase, \ce{V2Cu3H6O11} ($C2$, \#10), differs from Volborthite only by slight lattice distortions, and our calculated exchange ratios closely reproduce those of Volborthite, suggesting a comparable QSL-like behavior, although no magnetic measurements have yet been reported. Similarly, Kapellasite (\ce{ZnCu3H6(ClO3)2}, $P\bar{3}m1$, \#1)~\cite{mat_kapellasite2008,mat_kapellasite2012} yields exchange parameters in good agreement with Herbertsmithite and previous theoretical results~\cite{janson2008modified,jeschke2013first,kermarrec2014spin}. Collectively, these findings demonstrate the robustness of our screening workflow described in Sec.~II and its consistency with established theoretical expectations for kagome quantum spin liquids~\cite{Model_kolley2015,Model_liao2017,Model_changlani2018,Model_prelovvsek2020,Model_iqbal2021}.

Beyond these four benchmarks, five additional already-synthesized materials
also pass our screening and exhibit exchange hierarchies indicative of strong frustration:
\ce{Na2Ti3Cl8} ($R\bar{3}m$, \#7)\footnote{No long-range order down to 1.6,K \cite{mat_Na2Ti3Cl8_1,mat_Na2Ti3Cl8_2}; low-$T$ trimerization ($R\bar{3}m!\to!R3m$, 179–190,K) likely requires magnetoelastic couplings and non-Heisenberg interactions }, 
\ce{Na2Mn3Cl8} ($R\bar{3}m$, \#8)\footnote{Large moment $\sim$5\,$\mu_{\mathrm{B}}$/Mn; two incommensurate orders at 1.6/0.6\,K~\cite{mat_Na2Mn3Cl8_2}; weak $J_1$–$J_2$ with sizable dipolar (and possibly weak interlayer) couplings likely drive the ordering.}, 
\ce{Cs2ZrCu3F12} ($R\bar{3}m$, \#11)\footnote{Computed on high-symmetry $R\bar{3}m$ (Materials Project); experimentally undergoes $R\bar{3}m\!\to\!P2_1/m$ below $T_{\mathrm{N}}$ well above AFM order~\cite{mat_Cs2ZrCu3F12}, reconciling small $J_{2,3}$ from the undistorted phase.}, 
\ce{Cs2Cu3SnF12} ($R\bar{3}m$, \#12) \footnote{Similar to \ce{Cs2ZrCu3F12}, AFM appears after $R\bar{3}m\!\to\!P2_1/n$ above $T_{\mathrm{N}}$~\cite{mat_Cs2Cu3SnF12}; Materials Project lists the high-symmetry $R\bar{3}m$ used in our mapping.}, 
and \ce{BaV2Ni3(HO5)2} ($C2/m$, \#13) \footnote{Database assumes uniform Ni–Ni spacing; experiment reports inequivalent Ni–Ni distances~\cite{mat_BaV2Ni3(HO5)2}.} 
For more quantitative fidelity, however, one typically must go beyond the minimal $J_1$–$J_2$–$J_3$ Heisenberg mapping to include magnetoelastic couplings, non-Heisenberg interactions (e.g., biquadratic terms), dipolar interactions, and structural distortions.

We now focus on the unsynthesized candidates (blue labels in Fig.\ref{fig:mft_results}(b)). As described above, the final validation checks exclude \ce{Li2Mn3F8} ($P6_{3}mc$, \#4) and \ce{Li2Nb3Fe3O8} ($P6_{3}mc$, \#9). The remaining five materials exhibit several noteworthy features summarized below: \\[2mm]

{\bf \text{\ce{Li5VP2(O4F)2}} ($P3$, \#6) --- }
The detailed descriptions of its electronic structure, crystal symmetry and magnetic couplings are given in Fig.~5 and Table~V. A notable feature of this material is the small $J_1$ and the even smaller further-neighbor couplings. As shown in Fig.~\ref{fig:mft_results}(b) and Table~\ref{dft_table}, $J_2/J_1$ and $J_{3d}/J_1 \approx 0$, while $J_{3a}$ is also less than $5\%$ of $J_1$. This hierarchy is robust across an alternative DFT$+U$ functional and nine combinations of $(U,J_H)$ (Appendix, Table~V). It is therefore suggested that \ce{Li5VP2(O4F)2} is a promising candidate for the valence-bond type order proposed for spin-1 systems~\cite{changlani2015trimerized}. 
\\[2mm]


{\bf \text{\ce{Li2V3F8}} ($R\bar{3}m$, \#5) --- }
Our exchange mapping yields a dominant $J_1=0.922$~meV, while $J_2$, $J_{3\mathrm{a}}$ and $J_{3\mathrm{d}}$ are negligible (Fig.~\ref{fig:mft_results}(b) and Table~\ref{dft_table}). This exchange hierarchy is robust across alternative DFT$+U$ functionals and multiple $(U,J_H)$ choices (Appendix, Table~VI), placing \ce{Li2V3F8} in the $J_1$-only regime of the kagome Heisenberg model. In this sense, it closely resembles \ce{Li5VP2(O4F)2}.
Owing to its larger spin magnitude ($S = 3/2$), however, \ce{Li2V3F8} approaches the classical limit of the kagome antiferromagnet, where strong frustration gives rise to a classical spin liquid (CSL) manifold. The corresponding crystal and electronic structures are shown in Appendix, Fig.~6. 
\\[2mm]

{\bf \text{\ce{Li4Fe3WO8}} ($R\bar{3}m$, \#2) --- }
The calculated first-neighbor $J_1 = 0.653$ meV with the corresponding interatomic Fe--Fe distance is 3.09\AA ~ (Table~\ref{dft_table} and Fig.~7). A particularly notable feature of this compound is the exceptionally large across-hexagon third-neighbor coupling $J_{3d}$, which exceeds 45\% of $J_1$, and is larger than both $J_2$ and $J_{3a}$ (Fig.~\ref{fig:mft_results}(b) and Table~\ref{dft_table}). This unusually strong interaction is attributed to superexchange pathways mediated by the central \ce{W^{6+}} ions, located at the centers of the hexagonal plaquettes (Appendix, Fig.~7). 
This hierarchy is robust to the choice of DFT$+U$ functional and on-site parameters: across all calculations we obtain $J_{3d}/J_1 \ge 0.147$ (Appendix, Table~VII). A comparably large $J_{3d}$ competing with $J_1$ is known to stabilize exotic non-coplanar states, namely `cuboc-1' phase (Table~\ref{dft_table}). \\[2mm]

{\bf  \text{\ce{Li2MgCo3O8}} ($P2/m$, \#14) \& {\bf \ce{Li2MgCo3O8}} ($C2/m$, \#15) --- }
These two sister compounds are quite similar except for a slight distortion: the Co–Co distances agree to within $\sim$0.1~\AA, and the exchange couplings differ by less than 0.1~meV (Table~\ref{dft_table} and Appendix). A notable feature in this family is the magnitude of $J_2$ and $J_{3\mathrm{a}}$: their $J_2/J_1$ and $J_{3\mathrm{a}}/J_1$ values are largest in Fig.~\ref{fig:mft_results}(b).
This unusual interaction profile has important consequences. Even in the classical limit they point to a nontrivial magnetic order; based on the specific exchange hierarchy, we tentatively refer to the anticipated state as “cuboc-3”, distinct from the previously-reported cuboc-1 and cuboc-2 phases on the kagome lattice~\cite{messio2011lattice}.  Moreover, given the effective spin moment of $S \approx 1/2$, strong quantum fluctuations are expected to play a significant role, suggesting a possible link to chiral quantum spin liquid behavior. Broader implications for the resulting frustrated magnetism are further discussed below.
\\[2mm]


To summarize, mapping the exchange hierarchies of the five kagome-lattice candidates onto the classical $J_1$–$J_2$–$J_{3\mathrm{a}}$–$J_{3\mathrm{d}}$ phase space reveals a rich landscape of magnetic behavior (see below). The $J_1$-dominated compounds, such as \ce{Li5VP2(O4F)2} and \ce{Li2V3F8}, are expected to host CSL or valence-bond type order in the absence of magnetic Bragg peaks. In contrast, compounds such as \ce{Li4Fe3WO8} and \ce{Li2MgCo3O8} exhibit stronger further-neighbor couplings: the across-hexagon interaction $J_{3\mathrm{d}}$ stabilizes noncoplanar cuboc-1 order characterized by a $2{\times}2$ magnetic superstructure, while competition between the along-bond third-neighbor $J_{3\mathrm{a}}$ and $J_2$ interactions gives rise to a previously unreported cuboc-3 phase with an even larger real-space periodicity and a distinct reciprocal-space signature.

\subsection{Classical spin phase diagram on the kagome lattice}
\label{sec:class-Pd-kagome}
\begin{figure*} [hbt!]  
\includegraphics[width=0.9\textwidth]{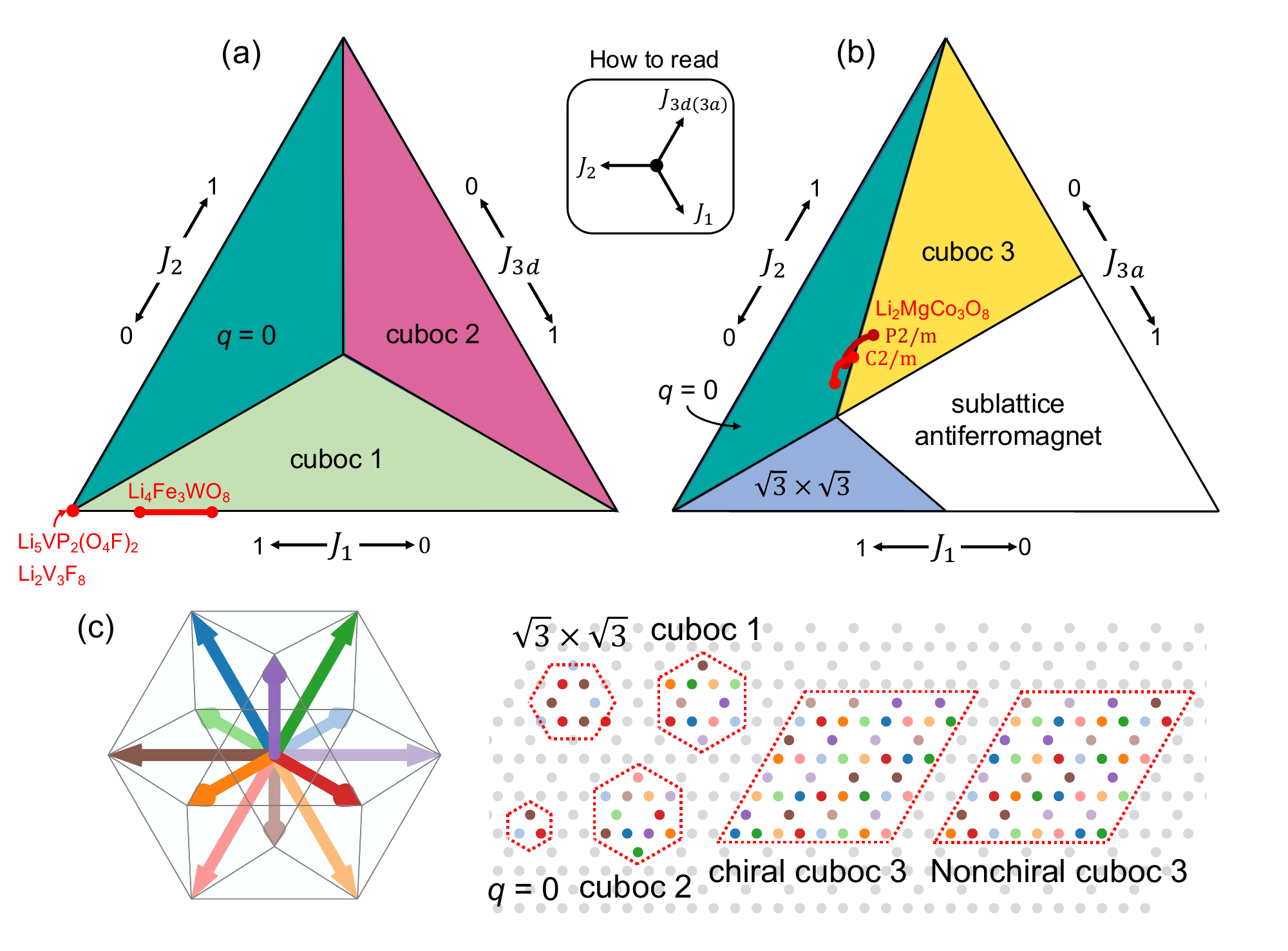}
\caption{Classical ternary phase diagrams for (a) \( J_{1}-J_{2}-J_{3d} \)  and (b) \( J_{1}-J_{2}-J_{3a} \) kagome Heisenberg models. Couplings are normalized such that \( J_{1} + J_{2} + J_{3a(3d)} = 1 \). The dashed region indicates the parameter regime of the material obtained from the DFT\(+U\) calculation. The inset illustrates how to read the ternary phase diagram. (c) Real-space spin configurations for each ordered phase. The spin colors correspond to the spin directions on the cuboctahedron shown on the left. 
}
\label{fig:phase_diagram}
\end{figure*}

Using the suggested workflow, we identified one triangular lattice and five kagome lattice candidates that have not yet been synthesized and are expected to host nontrivial magnetic phases. Notably, several kagome candidates exhibit sizable further-neighbor couplings in addition to the NN exchange. The kagome $J_{1}–J_{2}–J_{3d}$ Heisenberg model has been extensively studied in both classical and quantum regimes: classically it supports noncoplanar orders such as cuboc-1 and cuboc-2, while for $S=\frac{1}{2}$ it admits spin-liquid regimes (e.g., $U(1)$ and $Z_2$) and, for appreciable $J_2$ or $J_{3d}$, chiral spin-liquid phases \cite{Gong2014,PhysRevB.93.094437,Sun2024,PhysRevB.91.041124,Sun2024_1,PhysRevB.91.075112,PhysRevB.92.125122,PhysRevLett.108.207204}.  While the along-bond 3rd NN coupling $J_{3a}$ has also been studied, most prior work has concentrated on the special line $J_{2}=J_{3a}$ case~\cite{PhysRevResearch.5.L012025,PhysRevB.106.L140404,PhysRevLett.119.077207,PhysRevB.98.144446,PhysRevB.101.140403,PhysRevB.102.214424,PhysRevB.107.L020411,PhysRevB.94.235138,Li_2021}. 

In contrast, the generic case $J_{2} \neq J_{3a}$, relevant to our materials, remains comparatively unexplored.
Accordingly, we analyze the full $J_1–J_2–J_{3a(3d)}$ model relevant to our materials and show how a finite $J_{3a}$ reshapes the phase competition, potentially stabilizing new magnetic orders.
The Hamiltonian is given by
\begin{align}
\label{eq:hamiltonian}
    \mathcal{H}  
    &= J_{1}\sum_{\braket{i,j}}\vec{S}_{i}\cdot\vec{S}_{j} + J_{2}\sum_{\braket{\braket{i,j}}}\vec{S}_{i}\cdot\vec{S}_{j} \\ 
    &+ J_{3a(3d)}\sum_{\braket{\braket{\braket{i,j}}}} \vec{S}_{i}\cdot\vec{S}_{j}. \nonumber  
\end{align}  
Here, $J_{1}$, $J_{2}$, and $J_{3a}$ ($J_{3d}$) represent the antiferromagnetic Heisenberg interaction strengths for the first, second, and third NNs along the bonds (across the hexagonal plaquettes), respectively (see Fig.1(c)).

Fig.~\ref{fig:phase_diagram}(a) and (b) present the classical phase diagrams for the $J_{1}-J_{2}-J_{3d}$ and $J_{1}-J_{2}-J_{3a}$ models on the kagome lattice, using the Luttinger–Tisza method and the Landau–Lifshitz–Gilbert equation. In each phase diagram, we found various classical spin orderings: the $q=0$ phase, the sublattice antiferromagnetic phase, the $\sqrt{3} \times \sqrt{3}$ phase, and several types of cuboctahedral phases.\\[2mm]

\textit{$q=0$ phase} --- The first and second NN Heisenberg interactions on the kagome lattice can be rewritten in terms of the sum of spins on triangular plaquettes as
\begin{align}
    \mathcal{H}_{1,2} = J_{1,2} \sum_{ijk} \left( \vec{S}_i + \vec{S}_j + \vec{S}_k \right)^2 - (\text{const.}),
\end{align}
where $ijk$ denotes sets of three spins forming the small (for $J_{1}$) and large (for $J_{2}$) triangles of the kagome lattice. This form clearly indicates that energy is minimized when the spins on each triangle form a coplanar $120\degree$ configuration, satisfying $ \vec{S}_i + \vec{S}_j + \vec{S}_k = \vec{0} $. As both \( J_1 \) and \( J_2 \) terms independently favor this spin arrangement, the system naturally realizes a uniform \( 120\degree \) ordering on all small and large triangular plaquettes. The resulting spin configuration is characterized by a wave vector \( \vec{q} = \vec{0} \), where each magnetic unit cell coincides with the original unit cell (see Fig.~\ref{fig:phase_diagram}(c)). This ordered state is referred to as the \( q=0 \) phase.\\[1mm]

\textit{Sublattice Antiferromagnet} --- In the \( J_1-J_2-J_{3a} \) model, \( J_{3a} \) connects spins on the same sublattice across adjacent unit cells. In the limit of dominant \( J_{3a} \), the kagome lattice effectively decomposes into three decoupled NN antiferromagnetic Heisenberg models defined on square lattices. In this limit, the energy is minimized when each sublattice independently forms a Néel order, and the relative spin orientations between different sublattices become energetically irrelevant. 
This magnetic order pattern is referred to as the sublattice antiferromagnet phase.\\[1mm]


\textit{$\sqrt{3}\times\sqrt{3}$ phase} --- Among the degenerate classical ground states of the pure \( J_1 \) Heisenberg model on the kagome lattice, where the spins on each triangle satisfy the constraint \( \vec{S}_i + \vec{S}_j + \vec{S}_k = \vec{0} \), the inclusion of further-neighbor interactions can lift the degeneracy and select a specific spin configuration. In the case of dominant \( J_{3a} \) term compare to \( 
J_{2} \), which connects spins on the same sublattice across adjacent unit cells, the \( J_{3a} \) term favors a relative \( 120\degree \) spin arrangement among equivalent sublattices, effectively enlarging the magnetic unit cell. This results in a \( \sqrt{3} \times \sqrt{3} \) magnetic structure, as illustrated in Fig.~\ref{fig:phase_diagram}(c). 
We refer to this state as the \( \sqrt{3} \times \sqrt{3} \) phase due to the tripling of the original unit cell area.\\[1mm]

\textit{Cuboctahedron phases} ---
The classical phase diagrams of the \( J_{1}-J_{2}-J_{3d} \) and \( J_{1}-J_{2}-J_{3a} \) models reveal three types of cuboctahedral spin orderings: cuboc-1 and cuboc-2 in the $J_{1}-J_{2}-J_{3d}$ model~\cite{PhysRevB.92.220404,PhysRevB.93.094437,PhysRevB.83.184401,PhysRevB.91.075112}, and a novel cuboc-3 phase in the $J_{1}-J_{2}-J_{3a}$ model, as shown in Fig.~\ref{fig:phase_diagram}(c). While the cuboc-1 and cuboc-2 phases have been previously studied, the cuboc-3 phase emerges as a new classical spin configuration found in the current work. We provide a brief review cuboc-1 and cuboc-2 phases and then focus on a detailed characterization of the newly identified cuboc-3.\\[1mm]

\emph{Cuboc-1} : 
For aniferromagnetic $J_1$, the NN Heisenberg term enforces the local constraint
$\vec S_i+\vec S_j+\vec S_k=\vec 0$ on every primitive (up/down) triangle, which is
satisfied by coplanar $120^\circ$ configurations. Introducing an across-hexagon
third-neighbor coupling $J_{3d}$ selects, out of this manifold, a noncoplanar
twelve-sublattice state, the {cuboc-1} order. In this phase, the magnetic unit
cell enlarges to $2\times 2$ times the kagome unit cell (12 spins), with spin
directions pointing to the 12 vertices of a cuboctahedron (see
Fig.~\ref{fig:phase_diagram}(c)). While each small $J_1$ triangle remains
$120^\circ$ to minimize the NN energy, neighboring triangles are rotated out of a
common plane so that certain larger triangles acquire a finite scalar chirality,
$\vec S_i\!\cdot\!(\vec S_j\times\vec S_k)\neq 0$.\\[1mm]

\emph{Cuboc-2} :
Using a similar analogy to the cuboc-1 phase, \( J_{2} \) prefers \( 120\degree \) spin ordering to minimize the 2nd NN triangular plaquettes. When \( J_{3d} \) is introduced, each 2nd NN spin configuration preserves a coplanar \( 120\degree \) spin order, but the primitive triangular plaquettes become non-coplanar. Therefore, in contrast to the cuboc-1 phase, the primitive triangular plaquettes exhibit a non-zero scalar spin chirality, \( \vec{S}_{i} \cdot (\vec{S}_{j} \times \vec{S}_{k}) \neq 0 \) for $J_{2}-J_{3d}$ model. Nevertheless, the global spin configuration still places each spin direction on the vertices of a cuboctahedron, forming an enlarged $2 \times 2$ magnetic unit cell similar to cuboc-1 (see Fig.~\ref{fig:phase_diagram}(c)). To distinguish it from the cuboc-1 state, this phase is referred to as the cuboc-2 phase.\\[1mm]

\emph{Cuboc-3} :
In this study, we identify a novel type of cuboctahedral spin ordering in the \( J_1-J_2-J_{3a} \) model on the kagome lattice. This new ordering, which we refer to as the cuboc-3 phase, emerges from the competition between the \( J_{3a} \) and \( J_2 \) interactions. While the \( J_{3a} \) term favors Néel ordering within each sublattice, the \( J_2 \) interaction stabilizes coplanar \( 120\degree \) configurations on the large triangles of the kagome lattice. As a result, the spin configuration simultaneously satisfies the \( 120\degree \) coplanar condition on all large triangles and realizes antiferromagnetic order within each \( 2 \times 2 \) sublattice. This leads to an enlarged magnetic unit cell of \( 4 \times 4 \) relative to the original kagome unit cell (see Fig.~\ref{fig:phase_diagram}(c)). The spin directions align with the vertices of a cuboctahedron, but in a pattern distinct from the previously known cuboc-1 and cuboc-2 phases.

Interestingly, the cuboc-3 phase exhibits twofold degeneracy: a chiral and a non-chiral configuration. In the chiral cuboc-3 state, the scalar spin chirality \( \vec{S}_i \cdot (\vec{S}_j \times \vec{S}_k) \) is finite on each up and down small triangle plaquette. In contrast, the non-chiral configuration maintains coplanar spin order on all small and large triangles, resulting in vanishing scalar spin chirality, \( \vec{S}_i \cdot (\vec{S}_j \times \vec{S}_k) = 0 \), as shown in Fig.~\ref{fig:phase_diagram}(c). 

In the phase diagram shown in Fig.~\ref{fig:phase_diagram}(a) and (b), we indicate the parameter regimes corresponding to each kagome material candidate using red lines. It is noted that \ce{Li5VF2(O4F)2} and \ce{Li2V3F8} are located in the regime dominated by the NN interaction, effectively corresponding to the pure \( J_1 \) Heisenberg model (see  Fig.~\ref{fig:phase_diagram}(a)). Its classical ground states form an extensively degenerate manifold, known as the Coulomb spin liquid phase \cite{annurev:/content/journals/10.1146/annurev-conmatphys-070909-104138}.
These materials therefore provide the promising platforms to explore highly frustrated classical spin-liquid physics.

Next, \ce{Li4Fe3WO8} lies within the cuboc-1 stability region, having comparable $J_1$ and $J_{3d}$ (see  Fig.~\ref{fig:phase_diagram}(a)). This positioning in parameter space highlights \ce{Li4Fe3WO8} as a compelling platform for investigating kagome noncoplanar order and motivates the targeted experimental verification of the predicted cuboc-1 pattern.

Finally, \ce{Li2MgCo3O8} falls into a parameter regime where multiple phases, such as cuboc-3 and \( q=0 \), can potentially emerge (see  Fig.~\ref{fig:phase_diagram}(b)). The precise ground state depends sensitively on microscopic parameters, including the strength of Hund's coupling and on-site Coulomb interactions. In the quantum regime, only the special case with \( J_2 = J_{3a} \) has been studied so far \cite{PhysRevB.98.144446,Li_2021,PhysRevResearch.5.L012025}. This highlights \ce{Li2MgCo3O8} as not only an intriguing material candidate for realizing exotic magnetic orders but also as a promising platform for exploring quantum phases in the \( J_1-J_2-J_{3a} \) model. Further investigation of this model in the quantum regime thus represents an important direction for future work


\section{Computation Details}

DFT$+U$ calculations were performed using the OpenMX package \cite{openmx_package}, which employs a localized pseudoatomic orbital basis set. For the exchange–correlation functional, we adopted the generalized gradient approximation (GGA) \cite{pbe1992}, with a cutoff energy of 700 Ry and a ${\bf k}$-point mesh denser than 0.18 {\AA}$^{-1}$. As described above, we used the DFT$+U_{\rm eff}$ method \cite{dudarev1998, han2006n} at the first stage (the third screening step), with $U_{\rm eff}$ values taken from Ref.~\cite{moore_UJ2024}. Additional checks of the dependence on the choice of DFT$+U$ functional, double-counting formalism as well as on the Hubbard $U$ and Hund’s $J_H$, were performed at later stages (the sixth step). To verify the robustness of our results, additional computations were conducted with an alternative functional formalism~\cite{liechtenstein1995,cdft_ryee2018}, DFT spin-part control ~\cite{chen2015density,park2015density,chen2016spin,cdft_ryee2018} and varying the interaction parameters in a $\sim$10\% range. As mentioned above, two kagome candidate materials were excluded in this final validity check stage. For \ce{Li2Mn3F8}, charge-only DFT (cDFT) with the rotationally invariant Liechtenstein $+U+J_{\rm H}$ scheme yields a ferromagnetic $J_1$, in contrast to the spin-polarized GGA/LDA$+U$ mapping. Similar sensitivities to the functional choice (charge-only vs spin-polarized) have been reported for \ce{CrOCl}, \ce{CrOBr}, and \ce{LaMnO3}~\cite{jang2018charge,jang2021hund}. For \ce{Li2Nb3Fe3O8}, the Liechtenstein cDFT$+U+J_{\rm H}$ and Dudarev implementation agree only in the large-$U$ limit; at small $U$ they predict different ordered moments and opposite signs for the leading exchange coupling.

For MFT calculations, `Jx' code is used \cite{jxcode_yoon2018,yoon2020jx}. Full-band spin-dependent Kohn-Sham Hamiltonian taken from DFT$+U$ calculations serves as the input for MFT after the proper symmetrization process to improve the accuracy. The estimation of Eq.(\ref{Eq_Jij_all}) in the manuscript comes through 
\begin{equation} \label{Eq_green_DFT}
[{\mathbf{G}}_{{\bf{k}},ij}]_{l_1 l_2}^{\sigma} = \sum_{n}^{} \frac{ \ket{w_{l_1 ,{\bf{k}},i}^{\sigma}} \bra{w^{\sigma}_{l_2  ,{\bf{k}},j}} }{z-\epsilon_{\sigma, n,{\bf{k}}}   + i0^+},
\end{equation}
where $\sigma= \uparrow, \downarrow$ are spin indices, and the orbital $l_1$ and $l_2$ belong to the atomic site $i$ and $j$, respectively. $\ket{w^{\sigma}_{l_1 ,{\bf{k}},i}}$ refers to the local state. The final real-space value is obtained through Fourier transformation,
\begin{equation} \label{Eq_Jij_real2momentum}
J_{}^{} ( {\bf R}_{ij}  )  = \sum_{l_1 l_2} \tilde{J}_{}^{l_1 l_2} ( {\bf R}_{ij}  ) 
=  \sum_{l_1 l_2} \sum_{\bf{q}}^{} \tilde{J}^{l_1 l_2}_{ij} ({\bf{q} } ) ~ e^{-i {\bf R}_{ij} \cdot  {\bf q}}.
\end{equation}
At the earlier screening stage, magnetic couplings are estimated based on the ferromagnetic order for convenience, as it is expected to give a reasonable value for well-localized moments, which should be the case of our interest \cite{jxcode_yoon2018, yoon2020jx, szilva2023quantitative}. Eventually, this estimation is double-checked with the antiferromagnetic solution. We first take as many ${\bf q}$-points as in the ${\bf k}$-points mesh of the DFT calculations. After screening out, we double-check the candidate materials with a denser mesh whose largest grid spacing is less than 0.15 {\AA}$^{-1}$, and with optimized basis functions.

\section{Conclusion and Discussion}

We developed a targeted workflow for discovering frustrated magnets by integrating high-throughput first-principles calculations, magnetic-force theory, and spin-Hamiltonian analysis. Screening nearly 150,000 compounds, the framework identifies $3d$ kagome and triangular systems, reproduces known benchmarks, and reveals previously unexplored candidates with quantitatively resolved further-neighbor exchanges. By connecting large-scale electronic-structure data to model-level descriptions, it provides a predictive and physically transparent route from ab initio calculations to emergent magnetic behavior.

This approach demonstrates how high-throughput materials discovery can be made both selective and interpretable by focusing on motifs of magnetic frustration. The framework is readily extensible to include symmetry-based anisotropy diagnostics, uncertainty quantification, and machine-learning acceleration, and can be generalized to other frustrated lattices such as breathing or hyperkagome  networks and to tunable parameters including pressure, epitaxial strain, and chemical substitution.

Because several relevant exchanges are in the sub-meV range, small perturbations beyond the bilinear Heisenberg form such as Dzyaloshinskii–Moriya anisotropy, biquadratic and ring exchanges, and magnetoelastic couplings, may become decisive near phase boundaries. While our calculations confirm the robustness of the dominant exchange hierarchy, a comprehensive treatment of spin–orbit induced anisotropies and quantum-fluctuation effects represents a natural next step for future work.

\section{acknowledgements}
B. H. J. and H. S. K. contributed equally to this work. We thank Minu Kim, Jaewook Kim, and Jongmok Ok for valuable discussions on the possibility of experimental synthesis. B. H. J. and M. J. H. are grateful to Jiwon Kim for helpful support in data verification.
This work was supported by the National Research Foundation of Korea (NRF) grant funded by the Korean government (MSIT) (Grant Nos. RS-2025-02243032 and RS-2025-00559042).

\section{Data availability}
The data that support the findings of this study are available from the corresponding author upon reasonable request.



\section{Appendix}

\subsection{The detailed information of 32 triangular and 16 kagome materials (Table II and III)}


 \begin{table*} 
        \setlength{\tabcolsep}{8pt}
        \resizebox{\textwidth}{!}{
    \begin{tabular}{ccccccccccc} \hline\hline 
        number & mp-id & Formula & S.G. & m/TM ($\mu_{\mathrm{B}}$) & $J_{1}^{\mathrm{avg}}$ (meV) & $J_{1}^{\mathrm{STD}}$ & $J_{2}^{\mathrm{avg}}/J_{1}^{\mathrm{avg}}$ & $J_{3}^{\mathrm{avg}}/J_{1}^{\mathrm{avg}}$ \\
        \hline
        1  & mp-754490  & \ce{\textbf{Ni}HgO2}          & $R\bar{3}$   & $1.63$ & $3.06 $ & $0.04$ & $0.01$ & $0.22 $ \\
        2  & mp-6959    & \ce{Ga2\textbf{Ni}S4}         & $P\bar{3}m1$ & $1.37$ & $5.77 $ & $0.02$ & $0.01$ & $0.22 $ \\
        3  & mp-625939  & \ce{\textbf{Co}(HO)2}         & $C2$         & $2.74$ & $0.46 $ & $0.10$ & $0.00$ & $0.12 $ \\
        4  & mp-22877   & \ce{\textbf{V}Cl2}            & $P\bar{3}m1$ & $3.19$ & $0.13 $ & $0.00$ & $0.03$ & $0.13 $ \\
        5  & mp-2665409 & \ce{\textbf{Co}(HO)2}         & $C2/m$       & $2.74$ & $0.55 $ & $0.09$ & $0.00$ & $0.07 $ \\
        6  & mp-28233   & \ce{\textbf{Mn}Cl2}           & $R\bar{3}$   & $4.93$ & $0.04 $ & $0.00$ & $0.05$ & $0.12 $ \\
        7  & mp-1207060 & \ce{\textbf{Ti}I2}            & $P\bar{3}m1$ & $2.26$ & $1.02 $ & $0.01$ & $0.05$ & $0.06 $ \\
        8  & mp-27785   & \ce{\textbf{Ti}Br2}           & $P\bar{3}m1$ & $2.15$ & $0.44 $ & $0.01$ & $0.07$ & $0.07 $ \\
        9  & mp-28116   & \ce{\textbf{Ti}Cl2}           & $P\bar{3}m1$ & $2.07$ & $1.73 $ & $0.01$ & $0.03$ & $0.01 $ \\
        10 & mp-1228485 & \ce{Al3\textbf{Cr}O6}         & $R3$         & $3.16$ & $0.04 $ & $0.00$ & $0.04$ & $0.00 $ \\
        11 & mp-2210657 & \ce{Mg\textbf{Ti}O2}          & $R\bar{3}$   & $2.03$ & $15.13$ & $0.12$ & $0.01$ & $-0.04$ \\
        12 & mp-2222843 & \ce{KMg\textbf{Ni}IO6}        & $P312$       & $1.6 $ & $0.08 $ & $0.00$ & $0.15$ & $0.02 $ \\
        13 & mp-625671  & \ce{\textbf{Mn}(HO)2}         & $P1$         & $4.94$ & $0.12 $ & $0.01$ & $0.00$ & $0.05 $ \\
        14 & mp-2217116 & \ce{Mg\textbf{Mn}(CO3)2}      & $R\bar{3}$   & $4.96$ & $0.02 $ & $0.00$ & $0.00$ & $0.04 $ \\
        15 & mp-1078415 & \ce{Ca\textbf{Fe}SO}          & ${P6_{3}mc}$ & $3.76$ & $0.73 $ & $0.14$ & $0.01$ & $0.04 $ \\
        16 & mp-2210666 & \ce{Mg\textbf{Mn}O2}          & $R3m$        & $4.85$ & $0.64 $ & $0.00$ & $0.00$ & $0.02 $ \\
        17 & mp-755288  & \ce{Li\textbf{Fe}CuS2}        & $P3m1$       & $3.73$ & $1.26 $ & $0.21$ & $0.01$ & $0.02 $ \\
        18 & mp-1222418 & \ce{Na\textbf{Fe}CuS2}        & $P3m1$       & $3.72$ & $1.20 $ & $0.20$ & $0.01$ & $0.02 $ \\
        19 & mp-1208041 & \ce{Tl\textbf{Cr}(SO4)2}      & $P321$       & $3.23$ & $0.02 $ & $0.00$ & $0.01$ & $0.03 $ \\
        20 & mp-1105234 & \ce{Y\textbf{Cr}TeO6}         & $P\bar{3}1c$ & $3.13$ & $0.15 $ & $0.00$ & $0.00$ & $0.00 $ \\
        21 & mp-1177936 & \ce{Li2\textbf{Mn}(SeO3)2}    & $R\bar{3}$   & $4.93$ & $0.06 $ & $0.00$ & $0.00$ & $0.00 $ \\
        22 & mp-24100   & \ce{K\textbf{Cr}H18N6(ClO2)4} & $R\bar{3}$   & $3.47$ & $0.00 $ & $0.00$ & $0.00$ & $0.00 $ \\
        23 & mp-24099   & \ce{K\textbf{Cr}H18N6(ClO2)4} & $R\bar{3}m$  & $3.47$ & $0.00 $ & $0.00$ & $0.00$ & $0.00 $ \\
        24 & mp-1209246 & \ce{Rb\textbf{Cr}(SO4)2}      & $P321$       & $3.11$ & $0.22 $ & $0.00$ & $0.00$ & $0.00 $ \\
        25 & mp-1214516 & \ce{Ba3Nb2\textbf{Ni}O9}      & $P\bar{3}m1$ & $1.65$ & $0.17 $ & $0.00$ & $0.00$ & $0.00 $ \\
        26 & mp-684724  & \ce{\textbf{Fe}H4S2NO8}       & $R3$         & $4.15$ & $0.32 $ & $0.00$ & $0.00$ & $0.00 $ \\
        27 & mp-560628  & \ce{Cs\textbf{Fe}(SO4)2}      & $P\bar{3}$   & $4.14$ & $0.34 $ & $0.00$ & $0.00$ & $-0.01$ \\
        28 & mp-1218186 & \ce{Sr\textbf{Mn}TeO6}        & $P\bar{6}$   & $3.64$ & $0.18 $ & $0.00$ & $0.01$ & $0.01 $ \\
        29 & mp-1213686 & \ce{Cs\textbf{Cr}(MoO4)2}     & $P\bar{3}m1$ & $2.99$ & $0.17 $ & $0.00$ & $0.02$ & $0.01 $ \\
        30 & mp-1103452 & \ce{RbV2\textbf{Fe}(AgO4)2}   & $P\bar{3}$   & $4.06$ & $0.23 $ & $0.00$ & $0.01$ & $0.00 $ \\
        31 & mp-1103811 & \ce{KV2\textbf{Fe}(AgO4)2}    & $P\bar{3}$   & $4.05$ & $0.23 $ & $0.00$ & $0.01$ & $0.00 $ \\
        32 & mp-1227585 & \ce{CaTi8\textbf{Mn}(PO4)12}  & $R\bar{3}$   & $5.02$ & $0.00 $ & $0.00$ & $0.01$ & $0.00 $ \\
        \hline
    \end{tabular}\label{Table_Dudarev}
    }
\caption{The calculation results of 32 triangular materials listed in Fig.2(a): magnetic moments at transition-metal sites, $J_{1}^{\mathrm{avg}}$, the standard deviation of $J_1$, $J_{2}^{\mathrm{avg}}/J_{1}^{\mathrm{avg}}$, and  $J_{3}^{\mathrm{avg}}/J_{1}^{\mathrm{avg}}$. }
\label{supple_triangle_dudarev}
\end{table*}

 \begin{table*} 
        \setlength{\tabcolsep}{8pt}
        \resizebox{\textwidth}{!}{
    \begin{tabular}{ccccccccccc} \hline\hline 
        number & mp-id & Formula & S.G. & m/TM ($\mu_{\mathrm{B}}$) & $J_{1}^{\mathrm{avg}}$ (meV) & $J_{1}^{\mathrm{STD}}$ & $J_{2}^{\mathrm{avg}}/J_{1}^{\mathrm{avg}}$ & $J_{3\mathrm{a}}^{\mathrm{avg}}/J_{1}^{\mathrm{avg}}$ & $J_{3\mathrm{d}}^{\mathrm{avg}}/J_{1}^{\mathrm{avg}}$ \\
        \hline
        1  & mp-560161  & \ce{ZnCu3H6(ClO3)2}          &   $P\bar{3}m1$ & $0.48$ & $21.33 $ & $0.01$ & $-0.05$ & $ 0.02$ & $0.40 $ \\
        2  & mp-756528  & \ce{Li4\textbf{Fe}3WO8}      &   $R\bar{3}m$  & $3.80$ & $0.65 $ & $0.00$ & $ 0.01$ & $ 0.00$ & $0.45 $ \\
        3  & mp-1189352 & \ce{Zn\textbf{Cu}3H6(ClO3)2} &   $R\bar{3}m$  & $0.48$ & $116.49 $ & $0.00$ & $-0.08$ & $-0.04$ & $0.02 $ \\
        4  & mp-763793  & \ce{Li2\textbf{Mn}3F8}       &   $P6_{3}mc$   & $4.99$ & $0.08 $ & $0.04$ & $ 0.00$ & $ 0.01$ & $0.00 $ \\      
        5  & mp-778241  & \ce{Li2\textbf{V}3F8}        &   $R\bar{3}m$  & $3.11$ & $0.92 $ & $0.00$ & $ 0.00$ & $ 0.00$ & $0.00 $ \\      
        6  & mp-758318  & \ce{Li5\textbf{V}P2(O4F)2}   &   $P3$         & $2.21$ & $0.04 $ & $0.00$ & $ 0.00$ & $-0.04$ & $0.00 $ \\
        7  & mp-29474   & \ce{Na2\textbf{Ti}3Cl8}      &   $R\bar{3}m$  & $2.09$ & $0.10 $ & $0.00$ & $ 0.02$ & $ 0.08$ & $0.08 $ \\
        8  & mp-27126   & \ce{Na2\textbf{Mn}3Cl8}      &   $R\bar{3}m$  & $4.93$ & $0.03 $ & $0.00$ & $ 0.03$ & $ 0.16$ & $0.09 $ \\
        9  & mp-775011  & \ce{Li2Nb\textbf{Fe}3O8}     &   $P6_{3}mc$   & $4.08$ & $2.27 $ & $1.82$ & $ 0.03$ & $ 0.01$ & $0.00 $ \\
        10 & mp-1216869 & \ce{V2\textbf{Cu}3H6O11}     &   $C2$         & $0.56$ & $78.59$ & $15.67$ & $ 0.03$ & $ 0.00$ & $-0.01$ \\
        11 & mp-9202    & \ce{Cs2Zr\textbf{Cu}3F12}    &   $R\bar{3}m$  & $0.73$ & $26.42$ & $0.00$ & $ 0.05$ & $-0.04$ & $-0.01$ \\
        12 & mp-1106088 & \ce{Cs2\textbf{Cu}3SnF12}    &   $R\bar{3}m$  & $0.72$ & $21.81$ & $0.00$ & $ 0.05$ & $-0.03$ & $-0.01$ \\
        13 & mp-1205427 & \ce{BaV2\textbf{Ni}3(HO5)2}  &   $C2/m$       & $1.64$ & $2.39 $ & $0.06$ & $ 0.09$ & $ 0.08$ & $0.01 $ \\
        14 & mp-1277613 & \ce{Li2Mg\textbf{Co}3O8}     &   $P2/m$       & $1.28$ & $2.59 $ & $0.58$ & $ 0.11$ & $ 0.11$ & $0.01 $ \\
        15 & mp-1177928 & \ce{Li2Mg\textbf{Co}3O8}     &   $C2/m$       & $1.19$ & $2.77 $ & $0.54$ & $ 0.12$ & $ 0.10$ & $0.00 $ \\
        16 & mp-721062  & \ce{V2\textbf{Cu}3H6O11}     &   $C2/m$       & $0.55$ & $84.96$ & $19.96$ & $ 0.03$ & $ 0.05$ & $-0.10$ \\
        \hline
    \end{tabular}\label{Table_Dudarev}
    }
\caption{The calculation results of 16 kagome materials listed in Fig.2(b): magnetic moments at transition-metal sites, $J_{1}^{\mathrm{avg}}$, the standard deviation of $J_1$, $J_{2}^{\mathrm{avg}}/J_{1}^{\mathrm{avg}}$, $J_{3a}^{\mathrm{avg}}/J_{1}^{\mathrm{avg}}$, and $J_{3d}^{\mathrm{avg}}/J_{1}^{\mathrm{avg}}$. }
\label{supple_kagome_dudarev}
\end{table*}


\subsection{Crystal and electronic structure of six new materials (Fig.~4--9)}

\begin{figure*} 
\includegraphics[width=0.8\textwidth]{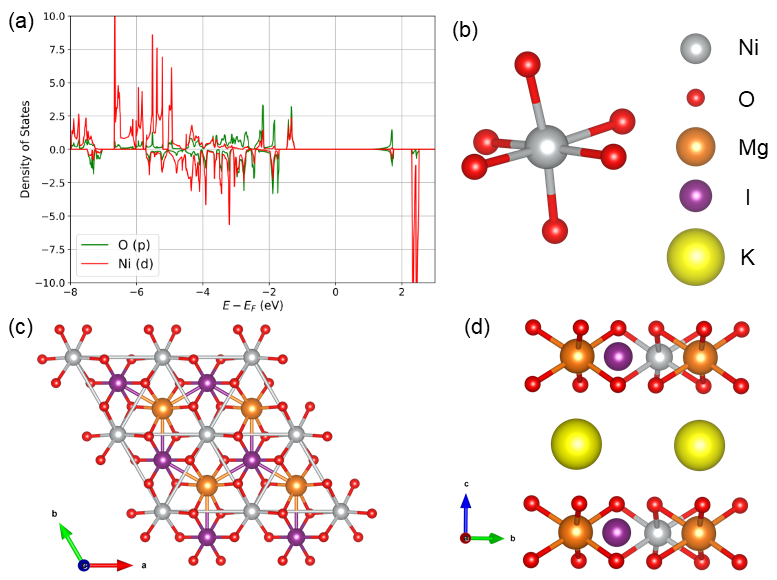}
\caption{(a) Projected density of states (PDOS) for \ce{KMgNiIO6} ($P312$). The red and green lines represent the Ni-$3d$ and O-$2p$ states (per atom), respectively. (b) The local atomic structure around transition-metal site (NiO$_6$). (c) Top and (d) side view of the crystal structure. The \ce{Ni}$^{2+}$ ions in the \ce{NiO6} octahedral environment adopt a $t_{2g}^6 e_g^2$ electronic configuration. Each triangular Ni layer is well isolated by non-magnetic \ce{K} layers, resulting in a quasi-two-dimensional magnetic lattice. The $C_3$ rotational symmetry along the $(001)$ axis and the $C_2$ symmetry along the $(210)$ direction within the triangular plane ensure that all NN exchange couplings ($J_1$) are equivalent. The NN Ni–Ni distance is 5.24\AA, and the magnetic superexchange pathway involves two distinct long-range bridges: \ce{Ni}–\ce{O}–\ce{I}–\ce{O}–\ce{Ni} and \ce{Ni}–\ce{O}–\ce{Mg}–\ce{O}–\ce{Ni}. These extended exchange paths give rise to a weak antiferromagnetic coupling with a small value of $J_1 = 0.079$ meV.}
\label{DOS_mp_2222843}
\end{figure*}

\begin{figure*} 
\includegraphics[width=0.8\textwidth]{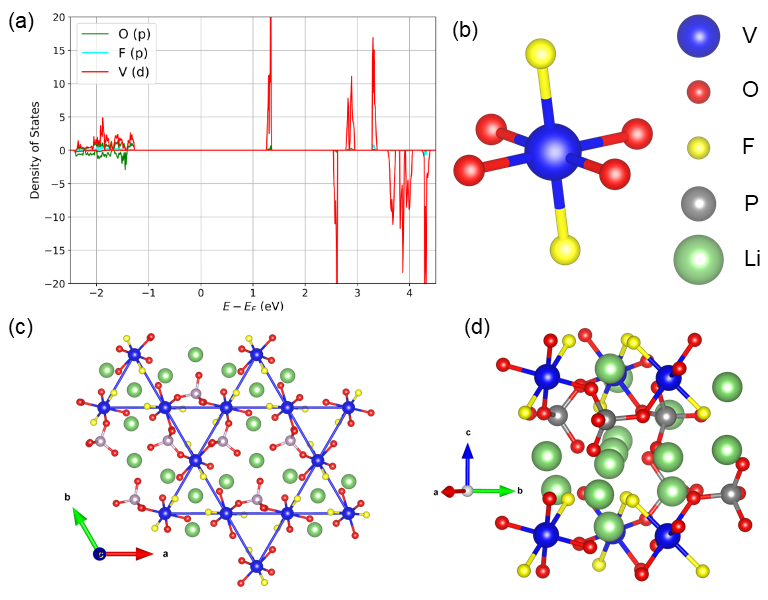}
\caption{(a) PDOS for \ce{Li5VP2(O4F)2}($P3$). The red, green and cyan lines represent the V-$3d$, O-$2p$ and F-$2p$ states (per atom), respectively. (b) The local atomic structure around transition-metal site (VO$_4$F$_2$). (c) Top and (d) side view of the crystal structure. The \ce{V^{3+}} ions, coordinated by \ce{O4F2} octahedra, nominally host two valence electrons in $t_{2g}$ orbitals. While $C_3$ rotational symmetry around the $(001)$ axis is preserved, the presence of inter-layer \ce{Li} ions breaks inversion symmetry, leading to a lowering of the overall lattice symmetry. As a result, the NN magnetic interactions split into two symmetry-inequivalent values, despite the underlying $C_3$ and translational symmetries. Consistent with this expectation, our MFT calculations yield two distinct NN couplings: $J_{1} = 0.035$ and $0.036$ meV. In the Supplementary Table.~\ref{supple_cDFT_mp_758318}, we report the average of these two values for simplicity. The long interatomic V–V distance of 5.05\AA, coupled with extended superexchange pathways (\ce{V}–\ce{O}–\ce{Li}–\ce{F}–\ce{V} and \ce{V}–\ce{O}–\ce{Li}–\ce{O}–\ce{V}) accounts for the relatively small magnitude of $J_1$.}
\label{DOS_mp_758318}
\end{figure*}

\begin{figure*} 
\includegraphics[width=0.8\textwidth]{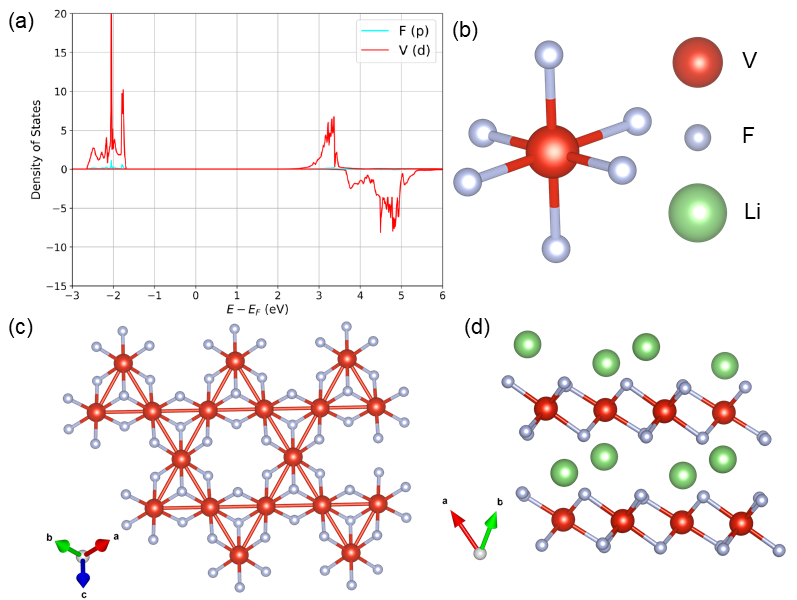}
\caption{(a) PDOS for \ce{Li2V3F8} ($R\bar3m$). The red and cyan lines represent the V-$3d$ and F-$2p$ states (per atom), respectively. (b) The local atomic structure around transition-metal site (VF$_6$). (c) Top and (d) side view of the crystal structure. The \ce{V^{2+}} ions at the centers of \ce{VF6} octahedra adopt a $t_{2g}^3$ configuration ($S=3/2$). The crystal symmetry keeps all NN bonds symmetry-equivalent, with a V–V distance of 3.14~\AA.}
\label{DOS_mp_778241}
\end{figure*}

\begin{figure*} 
\includegraphics[width=0.8\textwidth]{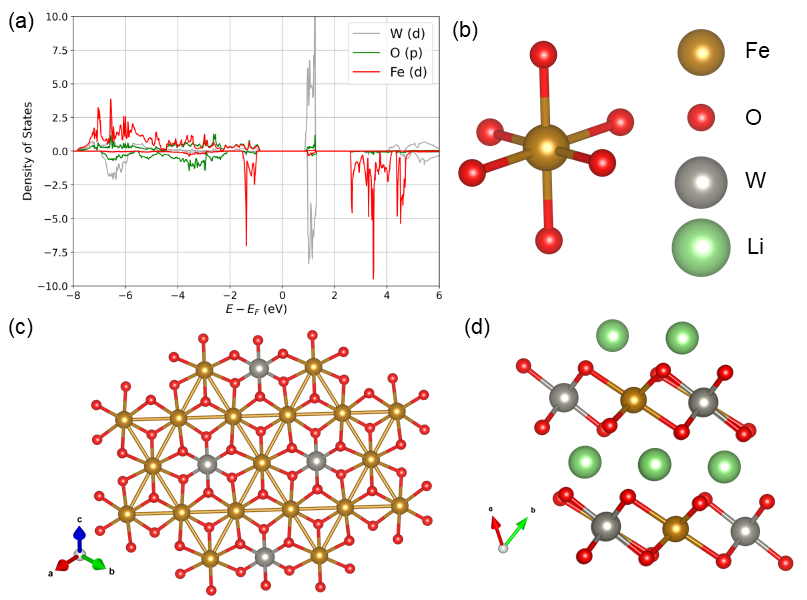}
\caption{(a) PDOS for \ce{Li4Fe3WO8} ($R\bar3m$). The red, green and gray lines represent the Fe-$3d$, O-$2p$ and W-$5d$ states (per atom), respectively. (b) The local atomic structure around transition-metal site (FeO$_6$). (c) Top and (d) side view of the crystal structure. The high-spin $t_{2g}^4 e_{g}^2$ configuration of \ce{Fe^{2+}} is captured by our calculations. }
\label{DOS_mp_756528}
\end{figure*}

\begin{figure*} 
\includegraphics[width=0.8\textwidth]{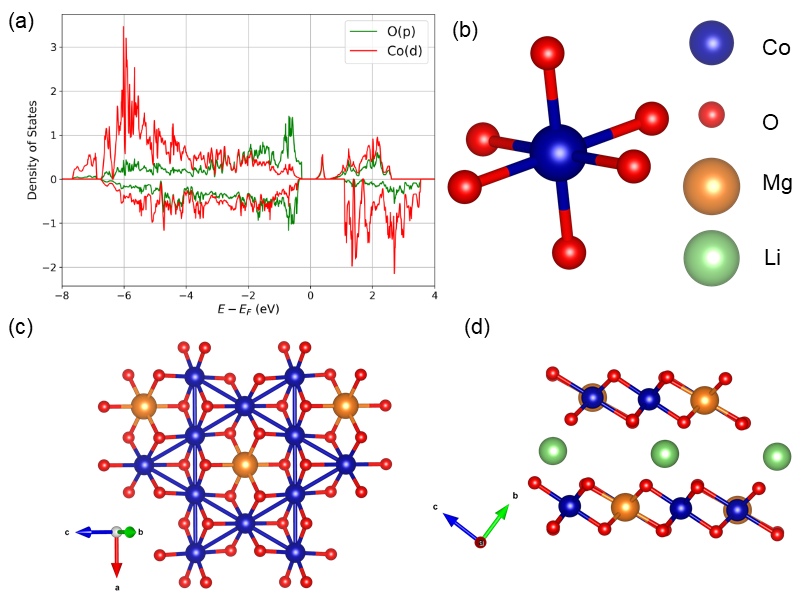}
\caption{(a) PDOS for \ce{Li2MgCo3O8} ($P2/m$). The red and green lines represent the Co-$3d$ and O-$2p$ states (per atom), respectively. (b) The local atomic structure around transition-metal site (CoO$_6$). (c) Top and (d) side view of the crystal structure. Being similar with \ce{Li2MgCo3O8} ($C2/m$), in $P2/m$ phase, the interlayer \ce{Li} ions and ABC stacking lower the full crystal symmetry to monoclinic, leaving only a $C_2$ axis. The partial breaking of $C_3$ symmetry, together with slight structural distortions, splits the NN exchange into symmetry-inequivalent paths, with a maximum deviation of 0.238 and 0.203~meV in $P2/m$ and $C2/m$, respectively. }
\label{DOS_mp_1277613}
\end{figure*}

\begin{figure*} 
\includegraphics[width=0.8\textwidth]{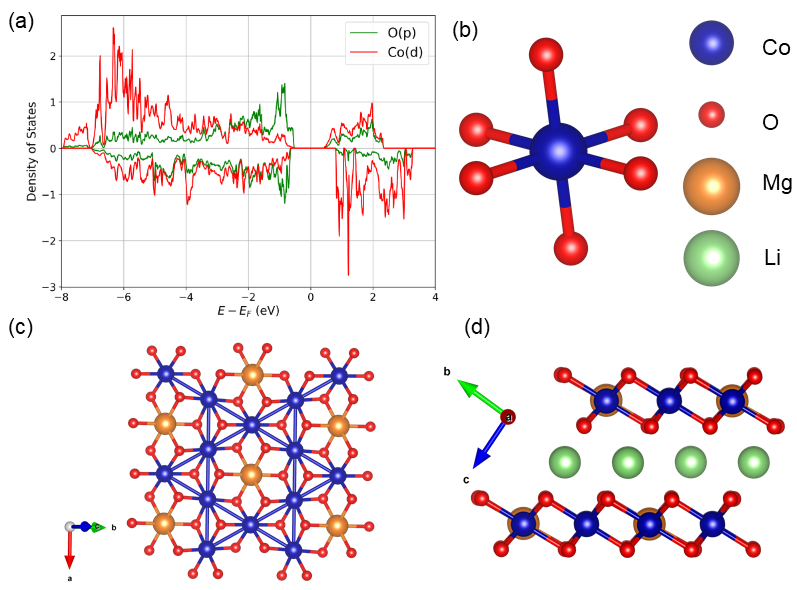}
\caption{(a) PDOS for \ce{Li2MgCo3O8} ($C2/m$). The red and green lines represent the Co-$3d$ and O-$2p$ states (per atom), respectively. (b) The local atomic structure around transition-metal site (CoO$_6$). (c) Top and (d) side view of the crystal structure. Our DFT$+U$ calculations capture their low-spin $t_{2g}^5$ configuration of \ce{Co^{4+}}. A single kagome-like layer comprises edge-sharing \ce{CoO6} octahedra, with \ce{Mg} located at the center of each hexagonal plaquette.
}
\label{DOS_mp_1177928}
\end{figure*}

\subsection{The exchange couplings for six new materials calculated within charge-DFT plus $U$ and $J_H$ functional (Table~IV--IX)}

\begin{table*} [b] 
\setlength{\tabcolsep}{8pt}
\centering
\begin{tabular}{|c|c|c|c|l|}
\hline
\ce{KMg\textbf{Ni}IO6} ($P312$) & $J_{\mathrm{H}}= 0.496$ eV & $J_{\mathrm{H}}= 0.682$ eV & $J_{\mathrm{H}}= 0.868$ eV & \\
\hline
\multirow{3}{*}{$U = 6.646$ eV}   & $0.120$ & $0.108$ & $0.097$ & $J_{1}$ (meV)           \\
                                  & $0.021 $ & $0.018 $ & $0.016 $ & $J_{2}$ (meV)   \\
                                  & $0.000 $ & $0.000 $ & $0.000 $ & $J_{3}$ (meV)   \\
\hline
\multirow{3}{*}{$U = 5.849$ eV}   & $0.143$ & $0.127$ & $0.114$ & $J_{1}$ (meV)           \\
                                  & $0.025 $ & $0.023 $ & $0.020 $ & $J_{2}$ (meV)   \\
                                  & $0.000 $ & $0.000 $ & $0.000 $ & $J_{3}$ (meV)  \\
\hline                
\multirow{3}{*}{$U = 5.052$ eV}   & $0.171 $ & $0.151 $ & $0.135$ & $J_{1}$ (meV)           \\
                                  & $0.031 $ & $0.028 $ & $0.025 $ & $J_{2}$ (meV)   \\
                                  & $0.000 $ & $0.000 $ & $0.000 $ & $J_{3}$ (meV)   \\
\hline                          

\end{tabular}
\caption{The calculated exchange interactions $J_{1,2,3}$ for \ce{KMgNiIO6} ($P312$) with varying $U$ and $J_{\mathrm{H}}$ values \cite{cdft_ryee2018}. In this validity check calculations, charge-DFT plus $U$ and $J_H$ functional originally suggested by Liechtenstein et al was used as described in Method part. While the absolute values of the exchange couplings vary by approximately a factor of two, depending on the specific functional and $U$–$J_H$ parameter choices, the relative magnitudes between $J_1$ and $J_2$  remain consistent, preserving the material’s position near the stripe–ordered regime in the phase diagram.
}
\label{supple_cDFT_mp_2222843}
\end{table*}


\begin{table*}  
\setlength{\tabcolsep}{8pt}
\centering
\begin{tabular}{|c|c|c|c|l|}
\hline
\ce{Li5\textbf{V}P2(O4F)2} ($P3$) & $J_{\mathrm{H}}= 0.437$ eV & $J_{\mathrm{H}}= 0.599$ eV & $J_{\mathrm{H}}= 0.761$ eV & \\
\hline
\multirow{4}{*}{$U = 4.594$ eV}   & $0.081$ & $0.070$ & $0.062$ & $J_{1}$ (meV)           \\
                                  & $0.000 $ & $0.000 $ & $0.000 $ & $J_{2}$ (meV)     \\
                                  & $0.000 $ & $0.000 $ & $0.000 $ & $J_{3\mathrm{a}}$ (meV) \\
                                  & $0.000 $ & $0.000 $ & $0.000 $ & $J_{3\mathrm{d}}$ (meV) \\
\hline
\multirow{4}{*}{$U = 4.123$ eV}   & $0.091$ & $0.079$ & $0.069$ & $J_{1}$ (meV)           \\
                                  & $0.000 $ & $0.000 $ & $0.000 $ & $J_{2}$ (meV)    \\
                                  & $0.000 $ & $0.000 $ & $0.000 $ & $J_{3\mathrm{a}}$ (meV) \\
                                  & $0.000 $ & $0.000 $ & $0.000 $ & $J_{3\mathrm{d}}$ (meV) \\
\hline                
\multirow{4}{*}{$U = 3.652$ eV}   & $0.103$ & $0.089$ & $0.077$ & $J_{1}$ (meV)       \\
                                  & $0.000 $ & $0.000 $ & $0.000 $ & $J_{2} $ (meV)    \\
                                  & $0.000 $ & $0.000 $ & $0.000 $ & $J_{3\mathrm{a}}$ (meV) \\
                                  & $0.000 $ & $0.000 $ & $0.000 $ & $J_{3\mathrm{d}}$ (meV) \\
\hline                          

\end{tabular}
\caption{The calculated exchange interactions $J_{1,2,3}$ for \ce{Li5VP2(O4F)2} ($P3$)  with varying $U$ and $J_{\mathrm{H}}$ values.}
\label{supple_cDFT_mp_758318}
\end{table*}


\begin{table*}  
\setlength{\tabcolsep}{8pt}
\centering
\begin{tabular}{|c|c|c|c|l|}
\hline
\ce{Li2\textbf{V}3F8} ($R\bar{3}m$) & $J_{\mathrm{H}}=0.437$ eV & $J_{\mathrm{H}}=0.599$ eV & $J_{\mathrm{H}}=0.761$ eV & \\
\hline
\multirow{4}{*}{$U = 4.594$ eV}   & $0.900$ & $0.655$ & $0.444$ & $J_{1}$ (meV)        \\
                                  & $-0.034$ & $-0.036$ & $-0.038$ & $J_{2}$ (meV)  \\
                                  & $-0.016$ & $-0.016$ & $-0.015$ & $J_{3\mathrm{a}}$ (meV) \\
                                  & $-0.006$ & $-0.007$ & $-0.007$ & $J_{3\mathrm{d}}$ (meV) \\
\hline
\multirow{4}{*}{$U = 4.123$ eV}   & $1.327$ & $1.042$ & $0.800$ & $J_{1}$ (meV)           \\
                                  & $-0.033$ & $-0.034$ & $-0.036$ & $J_{2}$ (meV)      \\
                                  & $-0.018$ & $-0.017$ & $-0.016$ & $J_{3\mathrm{a}}$ (meV) \\
                                  & $-0.005 $ & $-0.006$ & $-0.007$ & $J_{3\mathrm{d}}$ (meV) \\
\hline                
\multirow{4}{*}{$U = 3.652$ eV}   & $1.799$ & $1.456$ & $1.173$ & $J_{1}$ (meV)          \\
                                  & $-0.032$ & $-0.033$ & $-0.035$ & $J_{2}$ (meV)          \\
                                  & $-0.022$ & $-0.019$ & $-0.017$ & $J_{3\mathrm{a}}$ (meV) \\
                                  & $0.005 $ & $0.006 $ & $0.007 $ & $J_{3\mathrm{d}}$ (meV) \\
\hline                          

\end{tabular}
\caption{The calculated exchange interactions $J_{1,2,3}$ for \ce{Li2V3F8} ($R\bar{3}m$) with varying $U$ and $J_{\mathrm{H}}$ values.}
\label{supple_cDFT_mp_778241}
\end{table*}


\begin{table*}  
\setlength{\tabcolsep}{8pt}
\centering
\begin{tabular}{|c|c|c|c|l|}
\hline
\ce{Li2\textbf{Fe}3WO8} ($R\bar{3}m$) & $J_{\mathrm{H}}= 0$ eV & $J_{\mathrm{H}}= 0.279$ eV & $J_{\mathrm{H}}= 0.657$ eV & \\
\hline
\multirow{4}{*}{$U = 5.485$ eV}   & $2.386$  & $1.614$ & $1.152$ & $J_{1}$ (meV)           \\
                                  & $0.017 $  & $0.017 $ & $0.011 $ & $J_{2}$ (meV)      \\
                                  & $-0.047$  & $-0.031$ & $-0.030$ & $J_{3\mathrm{a}}$ (meV) \\
                                  & $0.858 $  & $0.428 $ & $0.194 $ & $J_{3\mathrm{d}}$ (meV) \\
\hline
\multirow{4}{*}{$U = 4.659$ eV}   & $3.372$ & $2.047$ & $1.374$ & $J_{1}$ (meV)           \\
                                  & $0.000 $  & $0.020 $ & $0.014 $ & $J_{2}$ (meV)        \\
                                  & $-0.098$  & $-0.038$ & $-0.029$ & $J_{3\mathrm{a}}$ (meV) \\
                                  & $1.273 $  & $0.555 $ & $0.222 $ & $J_{3\mathrm{d}}$ (meV) \\
\hline                
\multirow{4}{*}{$U = 3.833$ eV}   & $5.724$ & $2.773$ & $1.690$ & $J_{1}$ (meV)           \\
                                  & $-0.086$  & $0.021 $ & $0.017 $ & $J_{2}$ (meV)      \\
                                  & $-0.364$  & $-0.066$ & $-0.031$ & $J_{3\mathrm{a}}$ (meV) \\
                                  & $1.839 $  & $0.742 $ & $0.249 $ & $J_{3\mathrm{d}}$ (meV) \\
\hline                          

\end{tabular}
\caption{The calculated exchange interactions $J_{1,2,3}$ for \ce{Li2Fe3WO8} ($R\bar3m$) with varying $U$ and $J_{\mathrm{H}}$ values.}
\label{supple_cDFT_mp_756528}
\end{table*}



\begin{table*} 
\setlength{\tabcolsep}{8pt}
\centering
\begin{tabular}{|c|c|c|c|l|}
\hline
\ce{Li2Mg\textbf{Co}3O8} ($P2/m$) & $J_{\mathrm{H}}= 0.268$ eV & $J_{\mathrm{H}}= 0.560$ eV & $J_{\mathrm{H}}= 0.852$ eV & \\
\hline
\multirow{4}{*}{$U = 5.767$ eV}   & $5.262$ & $4.143$ & $3.296$ & $J_{1}$ (meV)           \\
                                  & $2.190 $ & $1.124 $ & $0.462 $ & $J_{2}$ (meV)          \\
                                  & $0.562 $ & $0.319 $ & $0.301 $ & $J_{3\mathrm{a}}$ (meV) \\
                                  & $-0.121$ & $-0.087$ & $-0.035$ & $J_{3\mathrm{d}}$ (meV) \\
\hline
\multirow{4}{*}{$U = 5.159$ eV}   & $6.439$ & $5.196$ & $4.123$ & $J_{1}$ (meV)           \\
                                  & $3.473 $ & $1.874 $ & $0.886 $ & $J_{2}$ (meV)     \\
                                  & $1.011 $ & $0.507 $ & $0.338 $ & $J_{3\mathrm{a}}$ (meV) \\
                                  & $-0.211$ & $-0.160$ & $-0.091$ & $J_{3\mathrm{d}}$ (meV) \\
\hline                
\multirow{4}{*}{$U = 4.551$ eV}   & $7.820$ & $6.477$ & $5.354$ & $J_{1}$ (meV)           \\
                                  & $5.138 $ & $2.911 $ & $1.506 $ & $J_{2}$ (meV)     \\
                                  & $1.746 $ & $0.837 $ & $0.452 $ & $J_{3\mathrm{a}}$ (meV) \\
                                  & $-0.323$ & $-0.284$ & $-0.187$ & $J_{3\mathrm{d}}$ (meV) \\
\hline                          

\end{tabular}
\caption{The calculated exchange interactions $J_{1,2,3}$ for \ce{Li2MgCo3O8} ($P2/m$) with varying $U$ and $J_{\mathrm{H}}$ values.}
\label{supple_cDFT_mp_1277613}
\end{table*}



\begin{table*} 
\setlength{\tabcolsep}{8pt}
\centering
\begin{tabular}{|c|c|c|c|l|}
\hline
\ce{Li2Mg\textbf{Co}3O8} ($C2/m$) & $J_{\mathrm{H}}= 0.268$ eV & $J_{\mathrm{H}}= 0.560$ eV & $J_{\mathrm{H}}= 0.852$ eV & \\
\hline
\multirow{4}{*}{$U = 5.767$ eV}   & $5.464$ & $4.258$ & $3.370$ & $J_{1}$ (meV)           \\
                                  & $2.079 $ & $1.112 $ & $0.497 $ & $J_{2}$ (meV)           \\
                                  & $0.476 $ & $0.267 $ & $0.253 $ & $J_{3\mathrm{a}}$ (meV) \\
                                  & $-0.114$ & $-0.085$ & $-0.026$ & $J_{3\mathrm{d}}$ (meV) \\
\hline
\multirow{4}{*}{$U = 5.159$ eV}   & $6.803$ & $ 5.363$ & $4.249$ & $J_{1}$ (meV)           \\
                                  & $3.266 $ & $1.806 $ & $0.885 $ & $J_{2}$ (meV)  \\
                                  & $0.882 $ & $0.435 $ & $0.278 $ & $J_{3\mathrm{a}}$ (meV) \\
                                  & $-0.192$ & $-0.156$ & $-0.086$ & $J_{3\mathrm{d}}$ (meV) \\
\hline                
\multirow{4}{*}{$U = 4.551$ eV}   & $8.299$ & $6.663$ & $5.376$ & $J_{1}$ (meV)           \\
                                  & $4.699 $ & $2.736 $ & $1.458 $ & $J_{2}$ (meV)      \\
                                  & $1.538 $ & $0.732 $ & $0.382 $ & $J_{3\mathrm{a}}$ (meV) \\
                                  & $-0.241$ & $-0.261$ & $-0.181$ & $J_{3\mathrm{d}}$ (meV) \\
\hline                          

\end{tabular}
\caption{The calculated exchange interactions $J_{1,2,3}$ for \ce{Li2MgCo3O8} ($C2/m$) with varying $U$ and $J_{\mathrm{H}}$ values.}
\label{supple_cDFT_mp_1177928}
\end{table*}


\end{document}